\documentclass[aps,prl,twocolumn,superscriptaddress,showpacs]{revtex4}

\usepackage{avant}
\usepackage[T1]{fontenc}

\usepackage{amsmath,amssymb,graphicx}
\usepackage{color,textcomp}

\usepackage[normalem]{ulem}

\newcommand\diff{\mathrm{d}}

\newcommand{\tauf}{\tau_{\text{f}}}
\newcommand{\taup}{\tau_{\text{p}}}
\newcommand{\tauw}{\tau_{\text{w}}}
\newcommand{\tauk}{\tau_{\text{k}}}
\newcommand{\massf}{m_{\text{f}}}
\newcommand{\massp}{m_{\text{p}}}

\begin{document}

\title{Anisotropic memory effects in confined colloidal diffusion }
\author{Sylvia Jeney}
\altaffiliation{Corresponding author. E-mail:
sylvia.jeney@epfl.ch}
\author{Branimir Luki{\'c}}

\affiliation{Institut de Physique de la Mati{\`e}re Complexe,
Ecole Polytechnique F{\'e}d{\'e}rale de Lausanne (EPFL), CH-1015
Lausanne, Switzerland}
\author{Jonas A. Kraus}
\author{Thomas Franosch}
\affiliation{Arnold Sommerfeld Center for Theoretical Physics (ASC) and Center for NanoScience (CeNS), Department of Physics,
 Ludwig-Maximilians-Universit{\"a}t M{\"u}nchen, Theresienstrasse 37, D-80333 M{\"u}nchen, Germany}
\author{L\'{a}szl{\'o} Forr{\'o}}
\affiliation{Institut de Physique de la Mati{\`e}re Complexe,
Ecole Polytechnique F{\'e}d{\'e}rale de Lausanne (EPFL), CH-1015
Lausanne, Switzerland}

\date{\today}

\begin{abstract}
The motion of an optically trapped sphere constrained by the vicinity of a wall is investigated at times where hydrodynamic memory is significant.
First, we quantify, in bulk, the influence of confinement arising from the trapping potential on the sphere's velocity autocorrelation function $C(t)$.
Next, we study the splitting of $C(t)$ into $C_\parallel(t)$ and $C_\perp(t)$, when the sphere is approached towards a surface. Thereby, we monitor the crossover from a slow $t^{-3/2}$ long-time tail, away from the wall, to a faster $t^{-5/2}$ decay, due to the subtle interplay between hydrodynamic backflow and wall effects. Finally, we discuss the resulting asymmetric time-dependent diffusion coefficients.

\end{abstract}



\maketitle

Understanding and controlling the transport of colloidal
microcarriers, such as membrane vesicles, through a fluid is one of
the main challenges in cell biology~\cite{Bareford2007}, and related lab-on-a-chip approaches. With the
miniaturization of such technologies, particularly microfluidics~\cite{Squires2005}, the colloid's motion is
increasingly confined, and the influence of boundaries, e.g. a
channel wall, becomes non-negligible. In particular, in a system as small as a
cell, many obstacles will alter the trajectory of a diffusing particle. Any deviation from its well-understood free Brownian motion, will give information on the particle's surroundings.
The  reduction of a colloid's mobility close to a wall,
also known as 'surface confinement', was already predicted by Lorentz in 1907~\cite{Lorentz1907,Happel1983}, and is expected to entail drastic effects on its time-dependent Brownian motion arising from the thermal fluctuations in the system.

Experimental evidences for this wall effect are comparably recent, and consist mostly in measuring changes in the diffusion coefficient $D$ of a micron-sized sphere approaching a  surface \cite{Faucheux1994,Carbajal-Tinoco2007,Schaeffer2007} or being confined between two walls \cite{Joly2006}. Measurements confirm that an interface increases the steady-state friction, hence the viscous drag on the
particle, slowing down its diffusion as the distance $h$ between the sphere's center and the surface is reduced. The motion becomes anisotropic in the directions parallel and perpendicular to the wall.
Conventional data acquisition and analysis rely on Lorentz's zero-frequency
approximation for the mobility, which only takes a viscous contribution from the fluid into account.
However, for a neutrally buoyant sphere, the presence of hydrodynamic memory due to momentum conservation in the fluid~\cite{Vladimirsky1945,Alder1967,Hinch1975} leads to it having a significantly delayed dynamic behavior, at time scales much larger than the particle's momentum relaxation time $\taup = 2a^2\rho_\text{p}/9 \eta$. This memory has
been observed in bulk for colloidal suspensions~\cite{Weitz1989,Zhu1992},
and, more recently, directly for a single microsphere~\cite{Atakhorrami2005,Lukic2005}. Its origin lies
in the back-flow that a spherical particle of density $\rho_{\text{p}}$ and radius $a$ creates in a fluid of viscosity $\eta$ and density
$\rho_{\text{f}}$. The long time diffusion of momentum in the viscous fluid leads to an
algebraic decay, the so-called long-time tail, in the velocity autocorrelation function (VACF) $C(t) =
\langle v(t) v(0) \rangle \simeq D \sqrt{\tauf/4\pi} t^{-3/2}$ of
the fluctuating sphere at times $t\gtrsim \tauf = \rho_{\text{f}} a^2/\eta$, the time
needed by the perturbed fluid flow field to diffuse over the
distance of one particle radius.
Obviously, the presence of a wall bounding the fluid will affect the fluid vortex
once it has encountered the wall, which will occur at times larger than its propagation
time, $\tauw = h^2 \rho_{\text{f}}/\eta$, from the particle to the boundary \cite{Felderhof2005}.

\begin{figure}
\includegraphics[width=\linewidth]{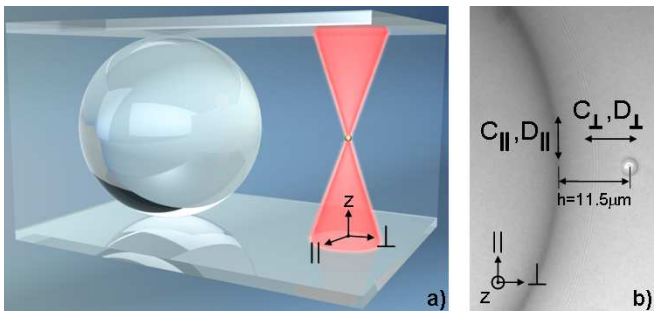}
\caption{\label{fig:scheme} (color)
(a) 3D lateral view of the experiment; spheres are drawn to scale. A silica particle of radius $a=$1.5\,\textmu{}m trapped by the laser focus is placed next to the surface of a
significantly larger silica sphere. This ~100\,\textmu{}m sphere is immobilized between
the two coverglass surfaces of the sample chamber.
(b) Optical image of the probing particle's position relative to the wall created by the big sphere.
The 3\,\textmu{}m probing particle was placed at a distance $h=$11.5\,\textmu{}m away
from the 100\,\textmu{}m sphere's surface and gradually approached. The velocity correlation functions, as well as the diffusion coefficients for the motion parallel and perpendicular to the wall are measured.}\end{figure}

In this Letter, we investigate the effect of hydrodynamic vortex diffusion on the nature of
the long-time tail as a function of the wall-particle distance $h$, by measuring the position fluctuations of a single spherical particle approaching a hard surface.
Weak optical trapping~\cite{Ashkin1986} is employed to position a silica sphere relative to
a plain glass surface, and, at the same time, to track the Brownian particle's trajectory
interferometrically~\cite{Gittes1998} with a precision better than 1\,nm and a time resolution
of 2\,\textmu{}s ~\cite{Lukic2005}.
The optical trap is created by focusing a 20$\times$-expanded Nd-YAG beam ($\lambda=1064$\,nm) by a 63$\times$
 water-immersion objective lens (NA$=1.2$). The thermal position fluctuations of the sphere in the weak trap
are measured with an InGaAs quadrant photodiode, amplified and digitized (12 bits). The position signal is
recorded during $t_s= 20$\,s with a sampling rate $f_s=500$\,kHz corresponding to $N=10^7$ data points.
The Brownian particle used here is a silica sphere of several micrometers diameter ($\rho_{\text{p}}=1.96$\,g/cm$^3$)
immersed in water ($\rho_{\text{f}}=1$\,g/cm$^3$, $\eta=10^{-3}$\,Pa$\cdot$s). It is gradually approached towards the surface of a
100\,\textmu{}m-sized sphere sandwiched between the two coverslides of a  fluid chamber (size $\approx 2$\,cm $\times$
0.5\,cm and thickness $\approx100$\,\textmu{}m). As illustrated in Fig.~\ref{fig:scheme},  the big sphere can be considered as a flat surface
on the  scale of
our Brownian sphere. Such a configuration  circumvents the draw-back of the
lower resolution along the optical axis $z$, intrinsic to optical trapping interferometry~\cite{Pralle1999},
thereby allowing measurements with comparable accuracy in both directions, parallel and perpendicular to the boundary.
The small silica spheres are suspended in water at a low enough concentration to allow trapping and observation of one isolated particle. The
sample is mounted onto a piezo-stage, and the 100\,\textmu{}m sphere can be positioned at a
distance $h$ relative to the trapped particle by moving the piezo-stage in all three dimensions with a precision of $\approx1$\,nm.

\begin{figure}
\includegraphics[width=\linewidth]{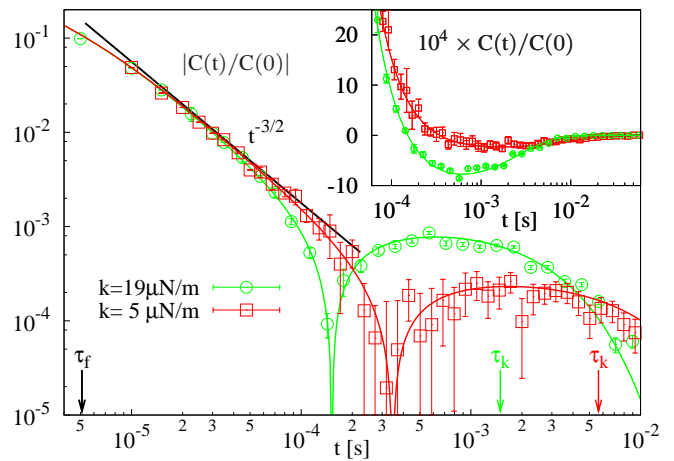}
\caption{\label{fig:LTT-bulk}(color) Log-log plot for the measured normalized velocity autocorrelation of
 a sphere ($a=$\,2.25\textmu{}m) held by an optical trapping potential with $k=19$\,\textmu{}N/m (green circles) and with
$k=5$\,\textmu{}N/m (red squares). The continuous red and green lines are fits to the model of Clercx and Schramm~\cite{Clercx1992}. The hydrodynamic long-time tail $C(t) \propto t^{-3/2}$ (black line) emerges as the spring constant is decreased.
Inset: Blow-up of the region of anti-correlations on a log-linear scale.}
\end{figure}

As a first test of the sensitivity of our set-up, we measure the VACF for a microsphere ($a=2.25$\,\textmu{}m, $\taup = 2.2$\,\textmu{}s, $\tauf = 5.1$\,\textmu{}s) far away from any boundaries ($h\geq$40\,\textmu{}m).
The optical trap is particularly well-suited for the experiment, as its harmonic potential results in a linear force $F=-k x(t)$, which only acts on the particle, but does not confine the surrounding fluid. The spring constant $k$ of the trap gives rise to a new time scale $\tauk = 6\pi \eta a/k$.
A model for the Brownian motion of a particle in a harmonic potential is provided by Clercx and Schramm~\cite{Clercx1992}, who went beyond a simple Stokes approximation by using
the time-dependant linearized Navier-Stokes equation to describe the fluid motion.
Fig.~\ref{fig:LTT-bulk} shows the measured normalized velocity autocorrelation function $C(t)/C(0)$ for $k=19$\,\textmu{}N/m and $k=5$\,\textmu{}N/m.
In order to compare data with the theoretical curves, the normalization $C(0)$ and $k$ are treated as fit parameters. As can be seen in Fig.~\ref{fig:LTT-bulk}, our data follow the theoretical prediction~\cite{Clercx1992} over the full time range, down to a noise level of $5\cdot 10^{-5}$.
As was shown earlier~\cite{Lukic2007}, an intermediate time window $\tauf \lesssim t \lesssim \tauk/20$ opens, where the particle's motion can still be observed, but is free from the influence of the trap and mainly dominated by the fluid's inertia.
Hence, in order to minimize influences from the trapping potential on the Brownian motion
and resolve hydrodynamic memory effects, we adjust $\tauk$ to be as long as possible. By decreasing the spring constant from $k=19$\,\textmu{}N/m to $k=5$\,\textmu{}N/m, we directly observe the emergence of the long-time anomaly $t^{-3/2}$ (Fig.~\ref{fig:LTT-bulk}) over
two decades in signal, which has not been previously reported in the literature. In an incompressible
liquid, the initial value of the VACF is determined by $C(0) = \langle v(0) v(0) \rangle = k_BT/(\frac{4\pi}{3}a^3(\rho_{\text{p}}+\frac{1}{2}\rho_{\text{f}}))$~\cite{Zwanzig1970}, leading to
\begin{equation}\label{eq:LTT-bulk}
\frac{C(t)}{C(0)} \simeq B \left(\frac{t}{\tauf}\right)^{-3/2}  \, , \qquad t \gtrsim \tauf \, ,
\end{equation}
where the amplitude $ B = ( \rho_{\text{p}}/\rho_{\text{f}} + 1/2)/ 9 \sqrt{\pi}$
depends only on the mass ratio. The value of the prefactor is of the order $B\sim 1/10$ for the colloidal system used here, and can be directly read off the data at $t=\tauf$ (Fig.~\ref{fig:LTT-bulk}).
At longer times, $t\gg\tauk$  ($t=22${}ms for $k=19$\,\textmu{}N/m and $t=96${}ms for $k=5$\,\textmu{}N/m),
the fluid's inertia is expected to generate a second zero in $C(t)$ followed by a fast $t^{-7/2}$-tail as $(15 B/4) (t/\tauf)^{-3/2} (t/\tauk)^{-2}$, which, however remains unobservable due to noise.

Next, we approach the Brownian sphere ($a = 1.5$\,\textmu{}m, $\taup =1$\,\textmu{}s, $\tauf = 2.25$\,\textmu{}s) towards
the boundary created by the 100\,\textmu{}m glass sphere (Fig.~\ref{fig:scheme}), and vary $h$ from 37.8\,\textmu{}m to 4.8\,\textmu{}m,
corresponding to a reduction in $\tauw$ from 1400\,\textmu{}s to 23\,\textmu{}s. The trap stiffness is minimized to $k$ $\approx$ 2\,\textmu{}N/m ($\tauk=$\,14\,ms) by lowering the incoming laser power. It is worth noting that decreasing $k$ degrades the signal-to-noise level, as can be seen in the increase of the error bars in Fig.~\ref{fig:LTT-bulk}. As already mentioned above, momentum from
the fluid is transferred to the wall at times $t\gtrsim \tauw$. The leading hydrodynamic tail $t^{-3/2}$ in $C(t)$ is
then canceled, and $C(t)$ splits into $C_\parallel(t)$ for the motion parallel and $C_\perp(t)$ for the motion perpendicular to the wall.
Recently, Felderhof~\cite{Felderhof2005} provided a full analytical solution for the motion in both directions.  He generalized the frequency-dependent admittance (\emph{i.e.} the frequency-dependent mobility)
for the unconstrained motion ${\cal Y}_0(\omega)$~\cite{Hinch1975}
to ${\cal Y}_\parallel(\omega)$ and ${\cal Y}_\perp(\omega)$ for a point particle moving close to a wall.
Relying on the fluctuation-dissipation theorem~\cite{Zwanzig1970}, he then calculated $C_\parallel(t)$ and $C_\perp(t)$
via a Fourier-back-transform from ${\cal Y}_\parallel(\omega)$ and ${\cal Y}_\perp(\omega)$.
For the parallel motion, a more rapid but still algebraic decay is predicted
\begin{equation}\label{eq:surface_1} \frac{C_\parallel(t)}{C_\parallel(t)} \simeq \frac{3 B}{2} \frac{\tauw}{\tauf}
\left(\frac{t}{\tauf} \right)^{-5/2} \, ,\end{equation}
for $t\gtrsim \tauw$, provided that the wall is not too close; $\tauw \gg \tauf, \taup$.
For the motion perpendicular to the wall, the long-time behavior, $t\gtrsim \tauw$, is predicted, to leading order in $a/h$, as
\begin{equation}\label{eq:surface_2}
\frac{C_\perp(t)}{C_\perp(0)} \simeq \frac{3 B}{2}  \left[ A_{z3}
\left( \frac{t}{\tauf} \right)^{-5/2} + \frac{\tauw^2}{4\tauf^2} \left( \frac{t}{\tauf} \right)^{-7/2}  \right] \, ,\end{equation}
where $A_{z3} = (2\rho_{\text{p}}/\rho_{\text{f}}-5)/9$.

To disentangle surface confinement from the trap constraint, we extend Felderhof's
theoretical approach~\cite{Felderhof2005}, by including the optical trapping force~\cite{Kraus2007}.
In bulk, the harmonic potential modifies ${\cal Y}_0(\omega)$~\cite{Hinch1975}
to ${\cal Y}_{\text{k}}(\omega) = [ {\cal Y}_0(\omega)^{-1}- k/\text{i} \omega]^{-1}$~\cite{Clercx1992}.
 Accordingly, in the presence of a wall, the admittances ${\cal Y}_{\parallel,\perp}(\omega)$ have to be modified to
$[{\cal Y}_{\parallel,\perp}(\omega)^{-1} - k/ \text{i} \omega]^{-1}$, and a suitable Fourier algorithm yields the VACFs in the entire time domain, which is fitted to our data.
\begin{figure}
\includegraphics[width=\linewidth]{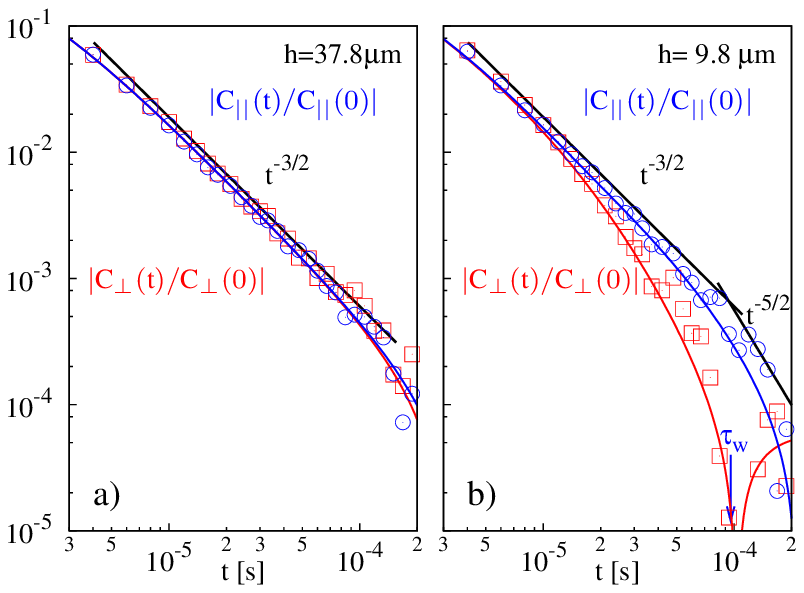}
\includegraphics[width=\linewidth]{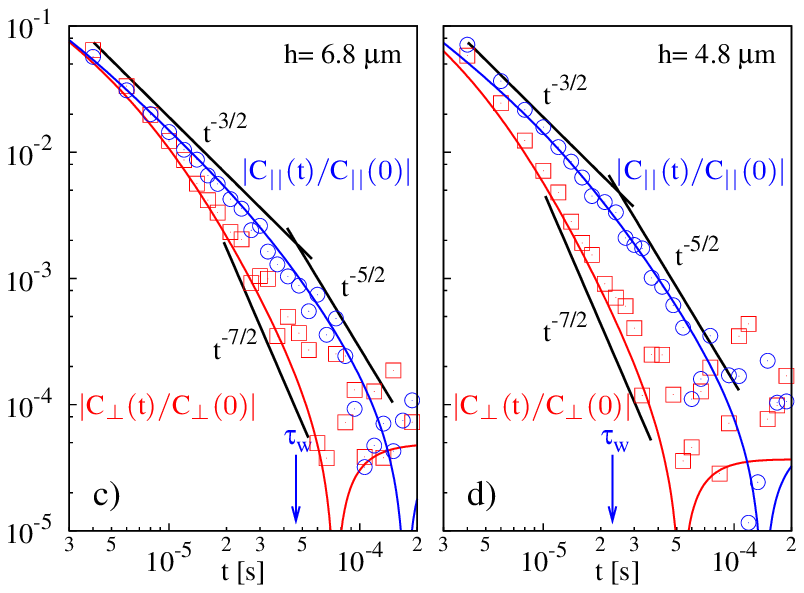}
\caption{\label{fig:LTT-wall}  Log-log plot of both normalized VACF, $C_\parallel(t)/C_\parallel(0)$ and $C_\perp(t)/C_\perp(0)$
for a sphere ($a=$1.5\,\textmu{}m, $\taup =1$\,\textmu{}s, $\tauf = 2.25$\,\textmu{}s) trapped in a weak optical potential ($k$ $\approx$ 2\textmu{}N/m, $\tauk=14$ ms ). The increasingly anisotropic VACF is measured at four distances from the wall ($h$=37.8, 9.8, 6.8 and 4.8\,\textmu{}m, corresponding to $\tauw$ = 1400, 96, 46, 23\,\textmu{}s, respectively).
The characteristic power-laws from Eq.~(\ref{eq:surface_1}) and Eq.~(\ref{eq:surface_2}) are represented by black lines as guides to the eye. The experimental data (squares and circles) are compared to the theory that includes hydrodynamic memory, wall effects, and harmonic restoring forces (continuous lines).}
\end{figure}

Figure \ref{fig:LTT-wall} shows the normalized $C_\parallel(t)$ and $C_\perp(t)$, in the time
range $\tauf<t<\tauk/20=0.7$ ms, at four different distances from the wall. Far away from the
boundary, at $h=$37.8\,\textmu{}m, we observe the same $t^{-3/2}$ power-law for both directions
(Fig.~\ref{fig:LTT-wall}a). However, as soon as the sphere reaches the proximity of the wall, and
 $\tauw$ falls into our window of observation, $C(t)$ splits  due to the hydrodynamic interaction
with the wall (Fig.~\ref{fig:LTT-wall}b, $h=9.8$\,\textmu{}m, and $\tauw = 96$\,\textmu{}s).
Our data show clearly that the anisotropy increases even more as $h$ is further decreased (Fig.~\ref{fig:LTT-wall}c and d).
The higher noise floor at smaller $h$ probably results from scattering of the highly divergent
trapping laser beam by the big sphere. In $C_\parallel(t)$, a transition from the free bulk
behavior at $h=$37.8\,\textmu{}m, characterized by the $t^{-3/2}$ power-law (Fig.~\ref{fig:LTT-wall}a),
to confined motion, with a steeper, $t^{-5/2}$ power-law arises in the data at
$t\simeq\tauw$. The effective amplitude in Eq.~(\ref{eq:surface_1}) is reduced by up to 30\% due to the trap.
The drop to negative values in $C_\parallel(t)$ arises from anticorrelations imposed by the harmonic trapping potential, and can only be captured by our extended theory~\cite{Kraus2007}.
As can be inferred from Figs.~\ref{fig:LTT-wall}, the theoretical curves describe the data down to the noise level.
For the perpendicular motion, the prefactor $A_{z3}$ of the leading $t^{-5/2}$-term in Eq.~(\ref{eq:surface_2})
depends on the relative densities, and is negative in our experiment, $A_{z3} = -0.12$.  Hence, for $k=0$, a sign change is expected at $t \simeq -\tauw^2/ 4\tauf A_{z3} = 1890, 8.53, 1.96,$ and $0.49$\,ms. According to Eq.~(\ref{eq:surface_2}), a positive tail $t^{-7/2}$ should dominate in the time window $\tauw \lesssim t \ll \tauw^2/\tauf$.
However, the sign change in $C_\perp(t)$ is observed to occur much earlier, even though the trapping potential is minimal. Nevertheless, for the two closest distances (Fig.~\ref{fig:LTT-wall}c and d), the steeper decay is compatible with the intermediate power law $t^{-7/2}$ predicted by Eq.~(\ref{eq:surface_2}), but with an amplitude reduced to 50\% by the trap. This influence from the trap as well as noise also obscure the crossover from the positive $t^{-7/2}$ to the negative $t^{-5/2}$ tail.

\begin{figure} \includegraphics[width=\linewidth]{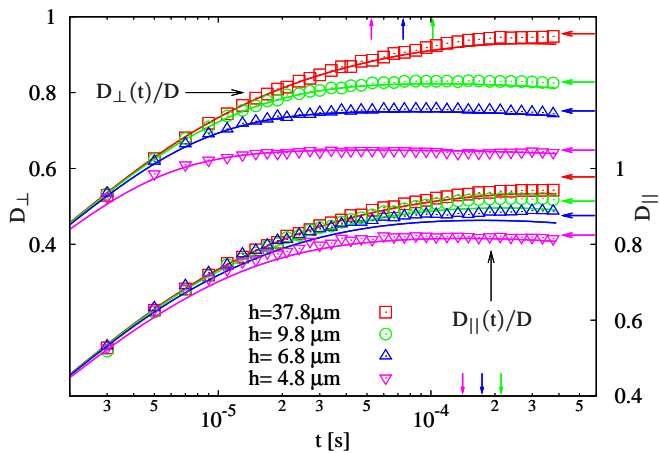}
\caption{\label{fig:Diffusion} Normalized time-dependent diffusion coefficients, $D_\parallel(t)/D$ and $D_\perp(t)/D$, for the direction parallel and perpendicular to the wall at the same distances $h$ as in Fig.~\ref{fig:LTT-wall}.
The arrows at the right correspond to the asymptotic values given by Lorentz's prediction in Eq.~(\ref{eq:Lorentz}).
The experimental data are represented by symbols, whereas the full lines correspond to the theoretical fits. The arrows at the top and the bottom indicate the respective time-points of the shallow maxima.}
\end{figure}

In general the fits in Fig.~\ref{fig:LTT-wall} are not very sensitive to the spring constant $k$. Therefore, we determine its value by comparing our data to the long-time behavior of the time-dependent diffusion coefficients, $D_{\parallel,\perp}(t)  = \int_0^t C_{\parallel,\perp}(t') \diff t'$, shown in Fig.~\ref{fig:Diffusion}.
As can be inferred from Fig.~\ref{fig:Diffusion}, the suppression of diffusion becomes anisotropic and follows the point-particle prediction by Lorentz~\cite{Lorentz1907},
\begin{equation}\label{eq:Lorentz}
D_\parallel=D[1-9a/16h]\, , \qquad D_\perp=D[1-9a/8h] \, ,
\end{equation} where $D = k_B T/6 \pi \eta a$ is the diffusion constant in bulk. The higher orders in $a/h$ are known to contribute less than $2\%$ for $h/a>3$~\cite{Happel1983}, and our data analysis suggests
that the point-particle limit is also valid for the time-dependent motion with similar accuracy. The observed zero-crossings in the VACF translate into shallow maxima in the diffusion coefficients.
They are still up to 5\% below the asymptotic values $D_\parallel$ and $D_\perp$, expected for $k=0$, which exemplifies that free Brownian motion is not attained in optical trapping of micron-sized spheres, even for times $t \lesssim \tauk$. At longer times the trap dominates and reduces the diffusion coefficients to zero (not shown).

In conclusion, we provide high precision experimental data relying on optical trapping interferometry which validate recent theoretical models. This approach gives access to the study of diffusion effects at a boundary, which are particularly relevant in diffusion-mediated interactions. During such processes, structures, like a large protein or membrane, interact with each other. This search for a potential interaction partner may be favored by the wall effect described here, as a fairly high mobility is preserved along the boundary, whereas diffusion away from the interaction partner is suppressed~\cite{Guigas2008}.


We thank E.-L. Florin, and F. H{\"o}fling for helpful discussions. SJ and BL acknowledge
support from the National Center of Competence in Research "Nanoscale Science", the Swiss National Science Foundation, and from the Gebert R{\"u}f Foundation. TF and JK acknowledge
support by the Nanosystems Initiative Munich (NIM).

\bibliographystyle{apsrev}

\end{document}